\documentclass[english,prl,twocolumn]{revtex4}
\usepackage{amsmath}
\usepackage{amssymb}
\usepackage{graphicx}
\usepackage{babel}

\begin{document}

\title{Growth of half-metallic CrO$_2$ nanostructures for superconducting spintronic applications }

\author{A. Singh$^{1}$, C. Jansen$^{1}$, K. Lahabi$^{1}$, \& J. Aarts$^{1}$}
\affiliation{$^{1}$Kamerlingh Onnes-Huygens Laboratory, Leiden University, P.O. Box 9504, 2300 RA Leiden, The
Netherlands.}

\begin{abstract}

Superconductor-ferromagnet (S-F) hybrids based on half-metallic ferromagnets, such as CrO$_2$2, are excellent
candidates for superconducting spintronic applications. This is primarily due to their fully spin polarized nature,
which produces significantly enhanced long-range triplet proximity effects.. However, reliable production of
CrO$_2$-based Josephson junctions (JJs) is challenging, mainly because of the difficulty to control the transparency of
the S/F interface. We have grown CrO$_2$ nanowires by chemical vapor deposition on TiO$_{2}$ substrates combined with
selective area growth in trenches defined with amorphous SiO$_x$. This allows us to create lateral JJs, with the
nanowire as the weak link. We show that the nature of the growth is highly anisotropic, and that the morphology of the
CrO$_2$ nanostructructures changes systematically during the growth process, depending on the width of the trench. The
detailed growth study enables us to synthesise multifaceted and highly homogeneous CrO$_2$ wires. These are utilized to
fabricate JJs with high S/F interface transparency, leading to large supercurrents. The well-defined geometry of the
device allows us to reliably estimate an exceptionally high critical current density J$_c$ = 10$^9$~Am$^{-2}$ over a
distance of 600~nm.
\end{abstract}
\date{\today}

\maketitle \selectlanguage{english}
Superconductor/Ferromagnet (S/F) hybrids have attracted great interest due their potential application in spintronic
devices to minimize the dissipation/Joule heating, which is the biggest challenge for current-driven processes at the
nanoscale~\cite{Eschrig_2016, Linder_2015,Buzdin_1}. The novelty of such hybrids lies in the generation of spin
polarized triplet Cooper pairs in the ferromagnet~\cite{berg_2001,Eschrig_2003,buzd_2007}. They are generated by
subjecting singlet Cooper pairs at the S/F interface to spin mixing and rotation~\cite{Eschrig_2008}, which is usually
achieved by introducing another thin ferromagnetic layer at the
interface~\cite{khaire_2010,birge_2012,Robinson59,Banerjee_2014}. Since triplet Cooper pairs are not affected by the
exchange field of the ferromagnet, they can survive over considerably long distances, thereby giving rise to
spin-polarized supercurrents. Fully spin polarized CrO$_{2}$ is an ideal ferromagnet for such applications. The spatial
extent of induced triplet correlations in CrO$_{2}$ is of the order of a micron~\cite{keiz_2006,anwar_2010}, while it
is limited to few tens of nm for standard ferromagnets (Co, Ni, Fe). Furthermore, the efficiency of the triplet
generation is sigficantly enhanced, as demonstrated in recent experiments in a spin valve
geometry~\cite{Singh_2015,stv_cambr}. Though large supercurrents have been observed in CrO$_{2}$
films~\cite{anwar_2012}, a serious bottleneck for further studies and applications of such Josephson junctions (JJs)
has been the poorly controlled interface transparency, together with ill-defined current paths in the full film.
Etching wire structures out of films proved no solution since the resist mask and the etch step are detrimental to the
surface and bring extra disorder in the material. To address this issue, we have made a detailed investigation into the
growth of CrO$_{2}$ nanowires by selective area chemical vapor deposition (CVD).

A salient feature of the CVD growth of CrO$_{2}$ is that it is only known to grow epitaxially on TiO$_{2}$ or sapphire
substrates. This is an advantage for selective area growth (SAG), whereby a patterned SiO$_x$ mask can be used to
define the geometry of the CrO$_{2}$ nanostructures. Previous reports on SAG of CrO$_{2}$ focussed on the morphology of
thick CrO$_{2}$ structures but did not give insight into the nucleation and growth kinetics at the early stage of
growth~\cite{SAG_1999,SAG_MFM_2007}. Here, we demonstrate the effect of mask confinement, lattice mismatch anisotropy,
and trench aspect ratio on the magnetic and structural morphology of CrO$_{2}$ nanostructures at different stages of
growth, covering a wide range of film thicknesses. A better understanding of SAG allows us to create high quality
uniform CrO${_2}$ nanowires with well-defined facets resulting in transparent interfaces. In addition, we show that the
magnetization state of the nanowire can be controlled by exploiting its shape anisotropy. This robust control over the
magnetization and transparency enables us to reproducibly fabricate CrO$_{2}$-based JJs with large critical current
densities. More generally, we have developed a device platform which will allow for a large variety of different wire,
contact and magnetic domain geometries.

\begin{figure*}[*t]
\includegraphics[width=0.8\textwidth]{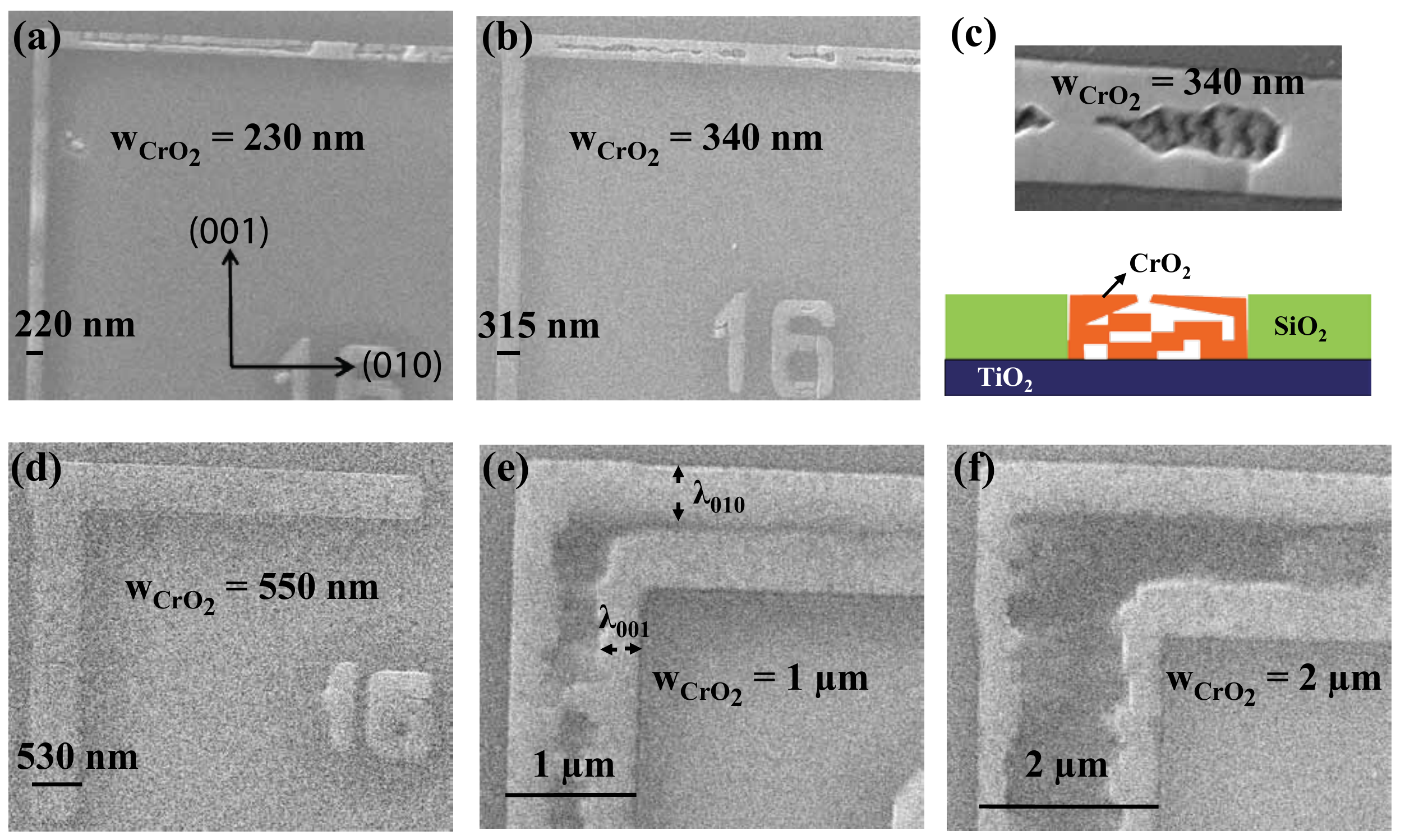}
\caption{Scanning electron micrographs of L-shaped nanowires along the indicated directions, grown for 20 minutes, with
trench widths (a) 230 nm, (b) 340 nm, (c) 340 nm, a close-up. The lower half schematically illustrates the growth mode
for trench widths (d) 550 nm, (e) 1 ${\mu}$m, and (f) 2 ${\mu}$m. The measured width of the CrO$_2$ bars is shown, and
found equal to the trench width. In (d) $\lambda$ is a measure for the surface diffusion length.}
\end{figure*}

{\bf Selective Area Growth} We start with a rutile (100) TiO$_{2}$ substrate which has a tetragonal surface net ($b$~=
0.459~nm, $c$~= 0.296~nm), on which CrO$_2$ ($b$~= 0.442~nm, $c$~= 0.292~nm) can be grown isostructurally. The lattice
mismatch is anisotropic, being -1.4~\% along the b-axis and -3.8~\% along the c-axis, respectively. This leads to
anisotropic growth rates, which is important for nanowire growth. Substrates were cleaned with organic solvents
(Acetone/IPA) followed by HF treatment, on which a 25~nm SiO$_{x}$ thin film was sputter-deposited. Using e-beam resist
and lithography, openings of different sizes and shapes were created in the resist along different crystallographic
directions. SiO$_{x}$ was selectively removed from the openings using RIE etching with a CF$_4$- and an O$_2$-plasma.
The etch time of SiO$_{x}$ is critical because underetching results in only partial removal of SiO$_{x}$ while
overetching damages the underlying TiO$_{2}$ substrate, which affects the film quality adversely. Next, CrO$_{2}$ was
grown selectively in the SiO$_{x}$ trenches (trench depth 25~nm, trench width $w_{trench}$) by CVD in a two-zone
furnace where the substrate temperature was kept at 395$^{\circ}$C while the precursor CrO$_{3}$ was heated to
260$^{\circ}$C in the presence of a flow of 100 sccm O$_{2}$ carrier gas. By monitoring the growth times (t${_{g}}$),
samples with varying thicknesses were prepared. Each sample consists of CrO$_{2}$ structures with different sizes,
shapes, aspect ratios and orientations. Scanning electron microscopy (SEM), Atomic force microscopy (AFM), and Magnetic
force microscopy (MFM) were used to examine the structural and magnetic morphology of
the CrO$_{2}$ nanostructures. \\
{\bf In-trench growth} Fig.~1a-e show the SEM images of L-shaped nanowires, grown for 20 minutes, with w$_{trench}$
varying from 230~nm to 2~$\mu$m. The time is chosen such that it fills the 25~nm deep trench. For the discussion, the
width range is divided into two regimes, Regime~I~: $w_{trench} \leq 2\lambda$, and Regime~II: $w_{trench} \geq
2\lambda$, where $\lambda$ is a characteristic length bearing resemblance to a surface diffusion length, which is
anisotropic due to the anisotropic lattice mismatch. For free films, growth results in the formation of rectangular
CrO$_{2}$ grains with their long axis along [001] and with an average size of $\sim$300 nm$\times$100 nm, which sets
the scale for $\lambda$. In the strongly confined structures shown in Fig.1a-b, nucleation and growth is clearly
anisotropic, with homogeneous wires forming along [001] and more inhomogeneous growth along [010]. This can be
understood from the better lattice match along [001] which also is the unconfined direction. Individual grains grow
faster and coalesce to form a continuous and homogeneous wire. Along [010] the wires grow in a peculiar fashion, namely
{\it from the sides} rather than from the bottom. Fig.~1c shows a close-up of the 340~nm trench where a smooth surface
is seen at the sides, and a hole in the middle where, at the bottom, growth has also started. Apparently,  CrO$_{2}$
islands nucleate preferably at the SiO$_{x}$/TiO$_{2}$ interface and grow laterally inward but also vertically upward.
This dynamically inhibits the supply of the reactants towards the bottom of the trench, but it does allow inward
lateral overgrowth starting higher up the wall. The result is a disordered structure with voids. This so-called
nanopendeo-epitaxy overgrowth (nPEO) is known from semiconductor growth, in particular for GaN, in the fabrication of
nanoairbridges and suspended structures of GaN with superior quality~\cite{PEO_nanolett,Nanoairbridge1,Pendeoepitaxi}.
\\
\begin{figure*}[*t]
\includegraphics[width=0.8\textwidth]{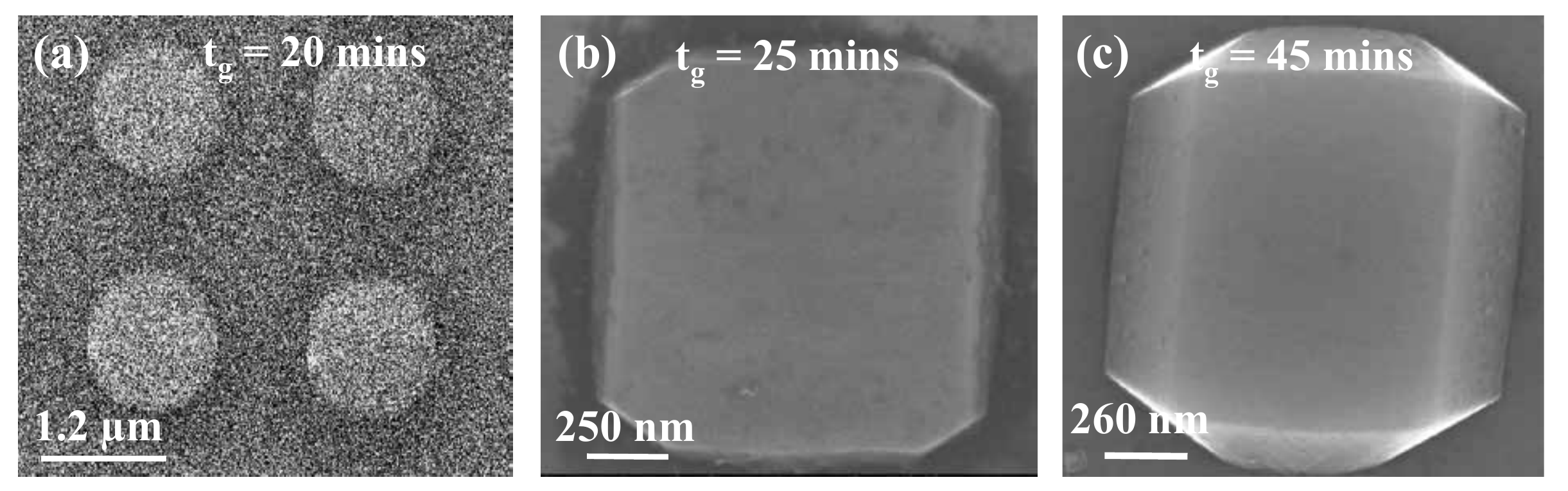}
\caption{Scanning electron micrographs of CrO$_{2}$ structures, grown in circular SiO$_{2}$ trenches, for (a) 20
minutes (filling the disk-shaped trench), (b) 25 minutes, and (c) 45 minutes. The latter two are overgrowth
structures.}
\end{figure*}
At $w_{trench}$ = 550 nm (Fig.1d) Regime II starts and the growth changes to a homogeneous filling of the
trench~\cite{SREP2015}. Islands nucleate along the walls and "quantum fortress"-like structures are formed irrespective
of the orientation of trenches~\cite{QuantumFortress}. Increasing $w_{trench}$ to 1 ${\mu}$m and 2 ${\mu}$m (Fig.1e,f)
the filling becomes inhomogeneous again, with the middle of the trench empty but now because the growth fronts
propagating from the walls do not reach the middle. We assume that the size of the CrO$_2$ strip along the walls is
determined by $\lambda$ and we observe $\lambda_{010} \sim$~450 nm and ${\lambda}_{010}$${\sim}$300 nm independent of
w${_{trench}}$. Note that these reults are quite different from earlier reports on SAG of CrO$_{2}$, where it was
assumed that CrO$_{2}$ nucleates and grows uniformly in the narrow trenches away from the SiO$_{x}$ side walls
~\cite{SAG_MFM_2007}. Our results present a very different picture, with important new aspects for CrO$_{2}$ nanowire
growth. \\
{\bf Lateral overgrowth} Next we focus on the lateral epitaxial overgrowth (LEO) above the SiO${_{x}}$ mask at higher
t$_g$. Fig.~2a-c shows the evolution of the structural morphology of CrO$_{2}$, grown in circular trenches, for
different t$_g$. For t$_g$ = 20 minutes, no overgrowth yet occurs, resulting in the formation of CrO$_{2}$ discs. With
further increase in t$_g$ (25 minutes), CrO$_{2}$ starts to overgrow laterally on SiO$_{x}$. Important to realize is
that there is no epitaxial relation with the amorphous SiO$_x$, so this is sideways growth starting from the structure
coming out of the trench. The development of (011) facets can be observed, assumedly to minimize the total surface
energy, although the facets formed at this stage are rough with step edges. For t$_g$ = 45 minutes, the CrO$_{2}$
islands exhibit fully developed facets with smooth surfaces. This type of facet evolution with growth time has been
reported recently for SAG
of GaN~\cite{SREP2015}. \\
In order to isolate the effect of lattice mismatch and size on the issue of CrO$_{2}$ overgrowth, we also prepared
rectangular 500 nm wide SiO$_{x}$ trenches oriented between the (010)- ($\theta=0^{\circ}$) and (001)-directions
($\theta=90^{\circ}$), at intervals of 15$^{\circ}$. The structural morphologies were characterized at different stages
(thicknesses) of growth. In Fig.~3a schematic of the LEO is outlined and Fig~3b,c show the SEM images of CrO$_{2}$
nanowires, with t$_g$ = 20 minutes, and 45 minutes, respectively. Clearly, there is no anisotropy in the lateral growth
for in the first 20 minutes of growth while it is highly anisotropic for t$_g$ = 45 minutes. As shown in Fig.~3c, the
nanowires exhibit jagged facets at the intermediate angles which is attributed to the higher growth rate along
(001)~\cite{SAG_1999}. The anisotropy of lateral overgrowth with SAG has been reported earlier for thick CrO$_{2}$
structures\cite{SAG_1999,SAG_MFM_2007}. However, from our data we can extract additional and quantitative information
on the anisotropy in lateral as well as vertical overgrowth over a range of sizes and thicknesses.
\begin{figure*}[*t]
\includegraphics[width=0.8\textwidth]{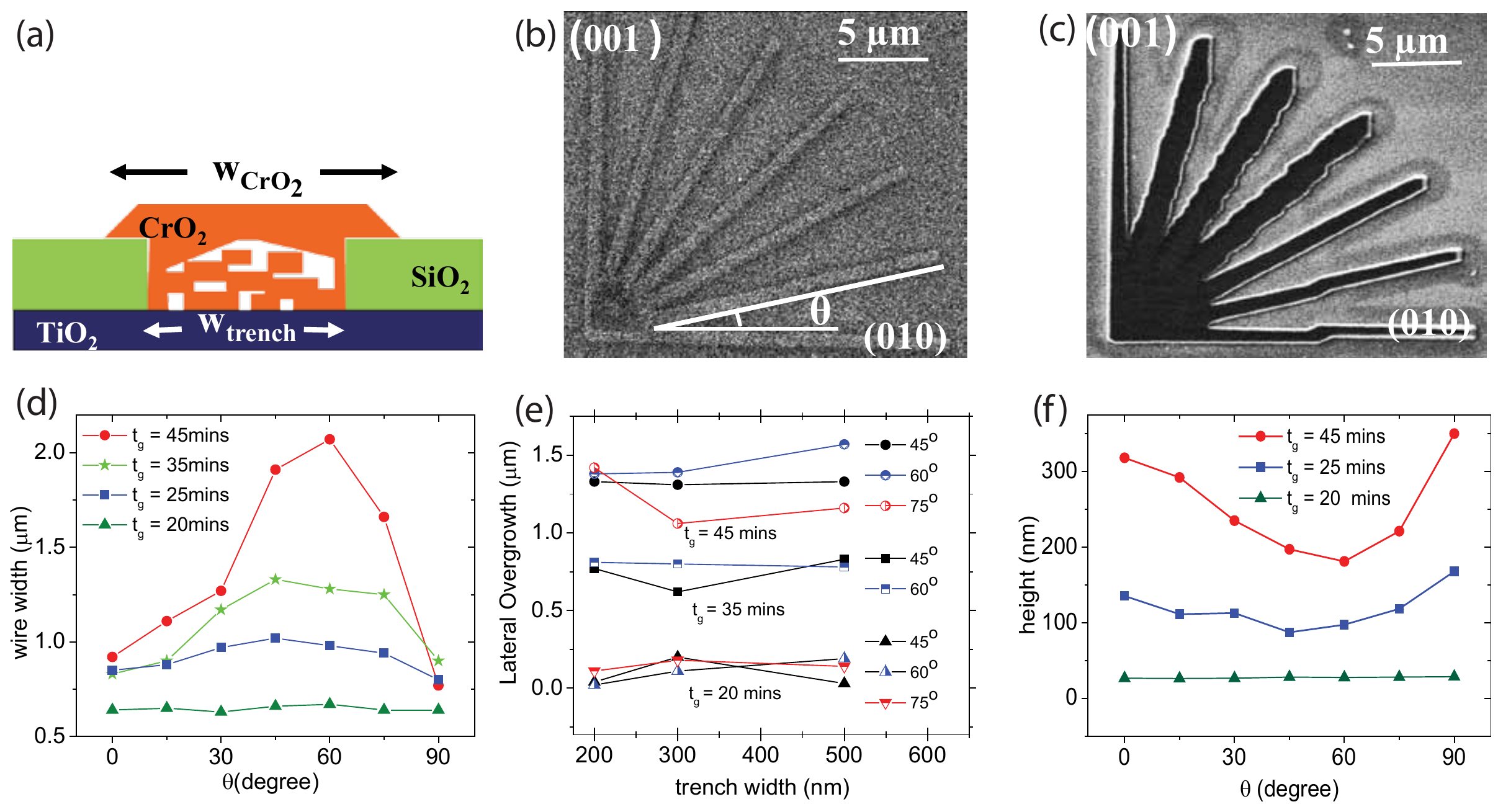}
\caption{ (a) Schematic of lateral epitaxial overgrowth of CrO$_{2}$ over SiO$_{2}$ for higher growth times. Scanning
electron micrographs of CrO$_{2}$ nanowires oriented between $\theta=0^{\circ}$ and $\theta=90^{\circ}$ at the
intervals of $15^{\circ}$ and grown in 500 nm wide trenches for (b) 20 minutes, and (c) 45 minutes. (d) Angular
dependence of width of CrO$_{2}$ nanowires (w$_{CrO_{2}}$). (e) lateral over growth (w$_{CrO_{2}}$-w$_{trench}$) as a
function of trench width (w$_{trench}$) along $\theta=45^{\circ}$, $60^{\circ}$, and $75^{\circ}$ for different growth
times. (f) Angular dependence of height of CrO$_{2}$ nanowires, grown in 500 nm wide SiO$_{x}$ trenches, for different
times.}
\end{figure*}
In Fig.~3d we plot the full width of the nanowire as measured from Fig.3c. This exhibits a maximum at $\theta=
60^{\circ}$ and becomes more pronounced with increasing $t_g$. The direction coincides with the diagonal of the
CrO$_{2}$ unit cell. In Fig.3e we plot the lateral overgrowth (defined as the difference between the total width of the
CrO$_{2}$ nanowire and the trench width) as function of $\theta$ for different trench widths up to 500 nm (Regime I).
The Figure shows that the lateral overgrowth hardly depends on trench width for the various angles, confirming that it
is mainly determined by the thickness and anisotropic strain. Finally, to study the anisotropy in vertical growth, AFM
was used to determine the thickness of the nanowires grown for different times up to $t_g$ = 45 minutes. Fig.~3f again
shows suggests that during the first 20 minutes the CrO$_{2}$  growth front just reaches the top of the SiO$_{2}$
trench and the thickness is isotropic. For higher growth times the vertical overgrowth becomes anisotropic, now with
the minimum along $\theta= 60^{\circ}$ : lateral overgrowth clearly goes at the expense of vertical overgrowth, with
the largest effect around 60$^o$ from the (010)-direction. This is in contrasts with the previous reports on SAG of
CrO$_{2}$
structures where vertical overgrowth was reported to be isotropic~\cite{SAG_1999}. \\
More formally, the correlation between the vertical and the lateral overgrowth, can be understood from the mass
transport mechanism i.e. absorption or desorption of adatoms on different facets of CrO$_{2}$ during the CVD growth.
The vertical growth depends on the impingent flux on top and on the sides of the growing surface and also on the
diffusion flux from the substrate to the top surface ~\cite{Growthkinetics_APL}. The size dependence of the vertical
growth rate is given by the following expression,
\begin{equation}
dH/dt= \gamma_{top}+ \gamma_{sw/D}+\gamma_{sub}/D{^2},
\end{equation}
where H is the height, D is the width of the structures, $\gamma_{top}$, $\gamma_{sw}$, and $\gamma_{sub}$ are the
constants related to surface impingement on top and  side wall and diffusion from the substrate, respectively. In this
simple picture, the maximum lateral overgrowth corresponds to largest D and hence lowest vertical overgrowth.
Furthermore, with change in orientation of the wire, $\gamma_{top}$ remains constant while the contribution from
$\gamma_{sw}$, and $\gamma_{sub}$ decreases due to the formation of jagged edges at the intermediate angles. Nonuniform
/ ragged edges prevent the surface diffusion of adatom on the side walls and also from the substrates, slowing
down the vertical  overgrowth. \\
{\bf Magnetic domains} Next we discuss the magnetic domain structures of our nanowires. Domain behavior has been
studied both in in early SAG investigations~\cite{SAG_MFM_2007} and in etched wires~\cite{Konig07,Bieler07} and it is
well known that for CrO$_2$ films on TiO$_2$ the [001] direction is the magnetic easy axis while the [010] direction is
the magnetic hard axis, as a consequence of the strain. Fig.4a shows the magnetic structure of wires of different size
along different crystallographic orientations as measured by using magnetic force microscopy (MFM). The wires oriented
along [001] are in a single domain state, as seen from the absence of any contrast in MFM image, and in the
cross-section shown in the lowest panel of Fig.~4b. When the wire is rotated away from the easy axis, striplike domains
start appearing in the wires to minimize the total magnetic energy. This can be seen by dark/bright contrast the MFM
micrograph (Fig.~4a) and as oscillations in the magnetic contrast (top and middle panel Fig.~5b). We observe that the
domain {\it size} is also direction dependent with a minimum along the [001] axis, which is the hard axis. The
wires therefore behave as expected. \\
\begin{figure*}[*t]
\includegraphics[width=0.8\textwidth]{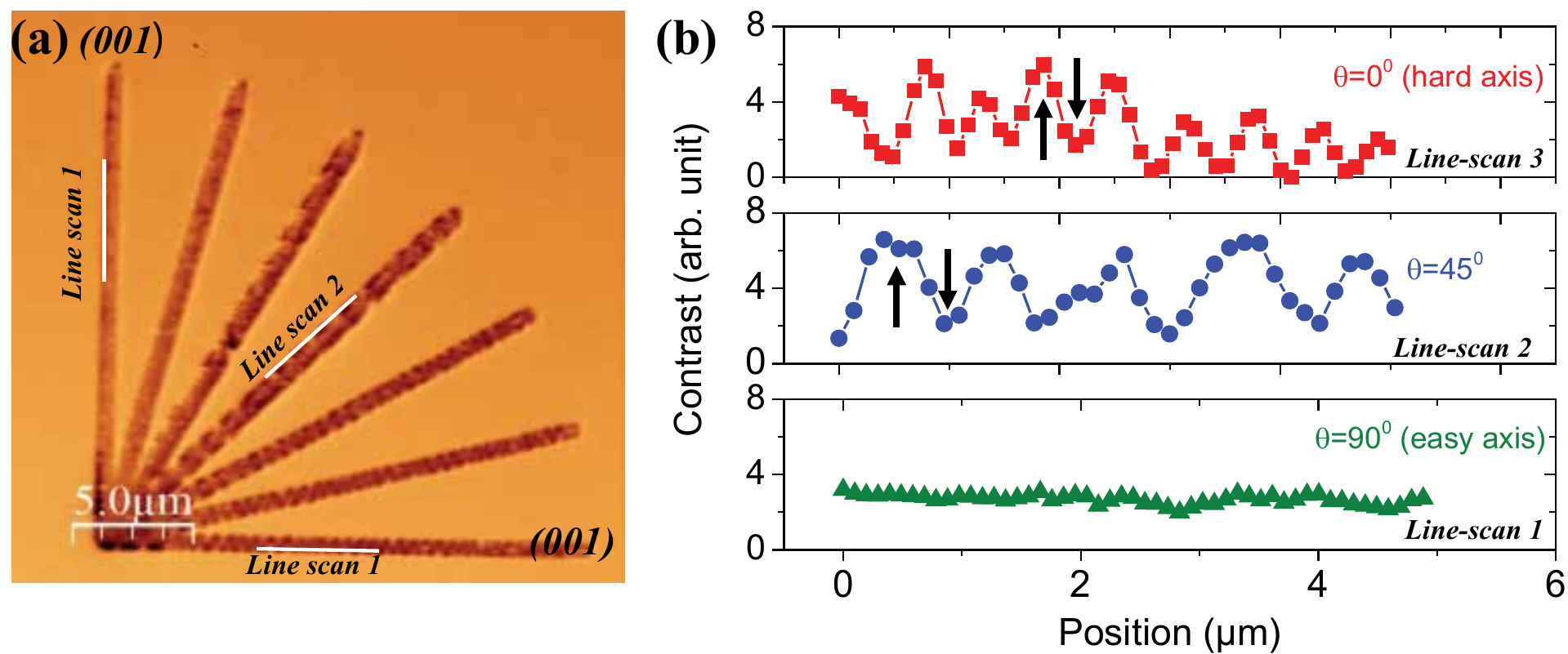}
\caption{(a) Magnetic force micrograph of CrO$_{2}$ nanowires grown in 500 nm wide trench, oriented between
$\theta=0^{\circ}$ and $\theta=90^{\circ}$ at the intervals of $15^{\circ}$. White along the edges of the wires
represent the line scan used to determine the phase magnetic phase contrast. (b) Magnetic phase contrast along the
edges of the wires  for $\theta=0^{\circ}$ (upper panel), $\theta=45^{\circ}$ (middle panel), and $\theta=90^{\circ}$
(lower panel). up and down arrow represent the opposite orientation (180$^{\circ}$) domains}
\end{figure*}
\noindent {\bf Supercurrents} As mentioned in the introduction, fully spin-polarized ferromagnetic nanowires are of
interest for superspintronics applications, and here we show that we can fabricate Josephson junctions with the CrO$_2$
nanowire as weak link. First we grow a nanowire along the easy axis [001] in order to have the magnetization along the
wire axis. Next we use lift-off and sputtering to deposit rectangularly shaped contacts of (Cu or Ag)(5 nm)/Ni(1.5
nm)/MoGe(125 nm) trilayer stacks, with their long axis perpendicular to CrO$_{2}$ wire, i.e. along the [010] direction
as shown in Fig.~5a. The distance between the contacts is typically 500~nm. The shape anisotropy aligns the
magnetization of the Ni layer in the contact stack along the crystal [010]) direction, which is very effective in
providing maximum non-collinearity at the interface, a key ingredient for triplet generation. In the trilayer stack we used either Cu or Ag to to prevent
exchange coupling between the CrO$_{2}$ and Ni layers. Amorphous MoGe is the superconductor used to induce the
proximity effect. Since at the surface CrO$_{2}$ inevitably reduces to the antiferromagnetic insulator Cr$_{2}$O$_{3}$
we clean it in-situ with a brief Ar-etch prior to the trilayer contact deposition. In this report, we present the
results obtained on two JJ devices, one with Cu and one with Ag as spacer
layer, named J1 and J2, respectively. The other parameters of J1 and J2 are provided in the table below. \\
\begin{table}[ht]
\caption{parameters of two Josephson Junction devices based on CrO$_{2}$ nanowires. The values for $R_N$ were
calculated using a specific resistivity of 6~$\mu\Omega$cm.} \centering
\begin{tabular}{c c c c c}
\hline
Device & width & thickness & junction length  & $R_N$  \\
\hline
J1 (Cu)&1 $\mu$ & 75 nm & 400 nm & 0.32 Ohm \\
J2 (Ag)& 500 nm & 150 nm & 600 nm& 0.48 Ohm \\ [1ex]
\hline
\end{tabular}
\label{table:nonlin}
\end{table}
Fig.5b shows a plot of resistance $R$ versus temperature ($T$) for J1 measured in a quasi-4-point geometry, meaning
that the interface resistance is part of the total normal state resistance. $R$ disappears around 6.5~K, but there is a
finite resistance tail down to 5~K, where R reaches 0 (c.q. the measurement limit). Below 5~K, supercurrents can be
measured. Fig.5c shows the current-voltage (I-V) characteristics of J1 at different temperatures from which the
critical current of the weak link (the proximized CrO$_{2}$) can be determined. Interestingly, despite its lower
junction length (see Table~I), J1 has a lower critical current density than J2, which is probably due to lower
interface transparency. Our earlier studies did not use Ag, but the result indicates that the use of Cu might lead to
degradation of the interface by the formation of some oxide. Fig.~5d displays the variation of critical current density
as a function temperature for J1 and J2, and as expected, I${_c}$ increases monotonically with the decrease in
temperature. The measured resistance of the device is dominated by the Cu layer in the leads, since the Cu specific
resistivity $\rho_{0,Cu}$ of the order of 5~$\mu \Omega$cm is much smaller than $\rho_{0,MoGe} \approx$ 200$ \mu
\Omega$cm. Taking a length of 70 $\mu$m per lead, a width of 2.5~$\mu$m and a thickness of 5~nm, we find 560~$\Omega$.
The normal resistance of the CrO$_2$ weak link we estimate using $\rho_{0,Cro2} \approx$ 6~$\mu \Omega$cm (the thin
films value for the easy axiss) to be 0.32~$\Omega$. This leaves about 20~$\Omega$ as the contact resistance. We note
in Fig.5b that the low-temperature R is well below 0.32~$\Omega$, and in Fig.5c that the voltage increase above I$_c$
starts with a small steep slope, but then settles to a slope which is close to 0.3~$\Omega$, quite as expected for the
CrO$_2$ weak link. Analyzing the junction behavior a bit further, in the long junction limit I${_c}$ is proportional to
$T^{3/2} exp(2{\pi}{k_B}T/E_{th})^{-1/2}$ where $E_{th}= \hbar D/L^2$ is the Thouless energy of the junction with
diffusion coefficient D and junction length L. Therefore, the slope of the plot of $ln(I_c)-{3/2}ln(T)$ vs $T^{1/2}$,
shown in the inset, can be used to estimate the Thouless energy $E_{Th}$ which is  11~$\mu$V. From this number we can
estimate D to be $2.7 \times 10^{-3}$ m$^2$/s, which is close to numbers obtained before. Using a Fermi velocity of
$2.2 \times 10^5$~m/s \cite{lewis97} it corresponds to a mean free path of 36~nm which suggests that our JJS are in
long (diffusive) junction limit.
\begin{figure*}[t]
\includegraphics[width=0.8\textwidth]{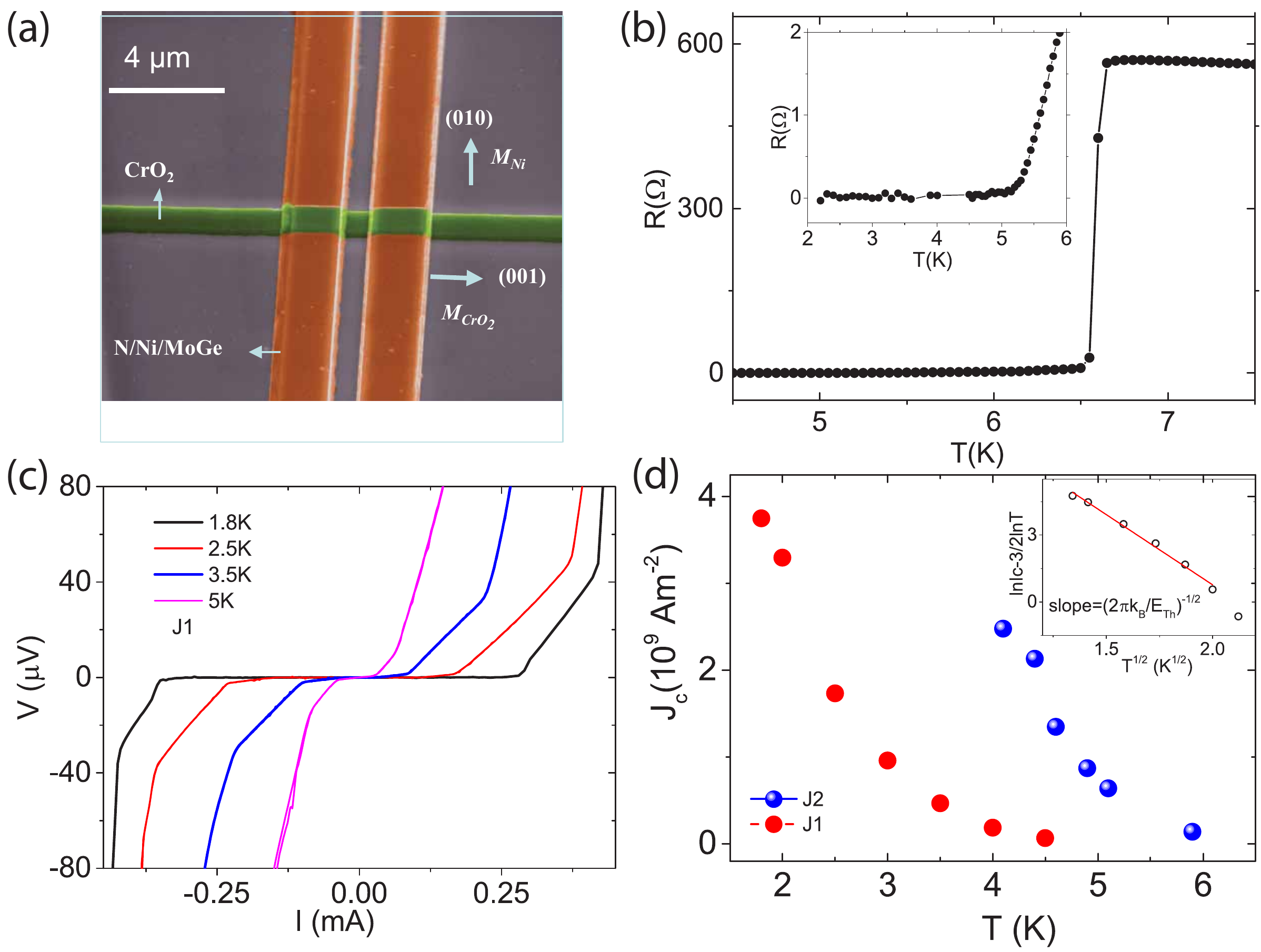}
\caption{(a). Scanning electron micrograph (false color) of a Josephson Junction on a CrO$_{2}$ nanowire (green). The
graph also shows crystal and magnetization directions. Orange contacts pads consist of trilayer (Cu or Ag)/Ni/MoGe. (b)
Resistance R versus temperature T for junction J1. The inset shows the behavior around zer resistance (c) Current I
versus voltage V characteristics of J1 at different temperatures. (d) Critical current density J$_c$ versus T; the
inset shows a linear fit to a plot of $ln(I_c)-{3/2}ln(T)$ vs $T^{1/2}$.}
\end{figure*}

{\bf Conclusions}  To conclude, we have made a detailed study of growth kinetics of halfmetallic ferromagnetic CrO$_2$
nanowires by selective area growth on TiO$_2$, in 25~nm deep trenches defined by a SiO$_x$ mask. At an early stage of
growth (t${_g}$=20 minutes), for trench width below 500 nm we find that thin homogeneous wires can be grown along
[001], while along [010] the growth is highly disordered, with voids, formed in a fashion called nanopendeo epitaxy.
Above a trench width of 500 nm, the trench is lined by CrO$_2$ strips formed along the edges with nonuniformity at the
center. By increasing the growth time, uniform CrO$_2$ structures are formed due to the lateral overgrowth over SiO$_x$
mask. Wires grown along the magnetic easy axis, which is the [001] direction, were contacted with stacks of Cu/Ni/MoGe
and Ag/Ni/MoGe, which allowed us to study triplet supercurrents. This upholds the earlier observation of very high
supercurrent densities of order $10^9$ A/m$^2$ at 4.2~K over a distance of 600~nm. CrO$_2$ wire based Josephson
junctions will make it now possible to characterize the interfaces fully, make reliable measurements of the dependence
of critical current on junction length, and study the influence of magnetic domains in the supercurrent path.

\vspace{0.5cm}

\noindent \textbf{ACKNOWLEDGEMENTS} Technical support from M.B.S. Hesselberth, D. Boltje, and A. F. Beker are
gratefully acknowledged.This work is part of the research programme of the Foundation for Fundamental Research on
Matter (FOM), which is part of the Netherlands Organisation for Scientific Research (NWO). The work was also supported
by the EU COST action MP1201 'NanoSC' and by a grant from the Leiden-Delft Consortium 'NanoFront'.


\bibliographystyle{unsrt}
\bibliography{Nanowires_Nanolett}

\end{document}